\documentstyle[preprint,prl,aps,eqsecnum]{revtex}

\def\inbar{\vrule height1.5ex width.4pt depth0pt}
\def\IC{\relax\hbox{\kern.25em$\inbar\kern-.3em{\rm C}$}}
\def\IP{\relax{\rm I\kern-.18em P}}
\def\IF{\relax{\rm I\kern-.18em F}}
\def\IZ{\relax\ifmmode\hbox{Z\kern-.4em Z}\else{Z\kern-.4em Z}\fi}
\def\IR{\relax{\rm I\kern-.18em R}}

\def\I1{\relax{\rm 1\kern-.35em 1}}

\begin{document}
{\tighten 
\title{\bf Compactifications of Type IIB Strings to Four Dimensions with
Non-trivial Classical Potential}
\author{Jeremy Michelson}
\address{e-mail: jeremy@cosmic1.physics.ucsb.edu \\
        Department of Physics \\ University of California \\
	Santa Barbara, California  93106}
\bigskip
\date{October 19, 1996}
\preprint{UCSB-TH-96-25, hep-th/9610151}
\maketitle
\begin{abstract}
Type IIB strings are compactified on a Calabi-Yau
three-fold.  When Calabi-Yau-valued
expectation values are given to the NS-NS and RR three-form field strengths,
the dilaton hypermultiplet becomes both electrically and magnetically
charged.  The resultant classical potential
is calculated, and minima are found.
At singular points
in the moduli space, such as Argyres-Douglas points,
supersymmetric
 minima are found.
A formula for the
classical potential in $N=2$ supergravity
is given which holds in the presence of both electric and magnetic charges.
\end{abstract}

} 

\section{Introduction} \label{intro}
It has been known for many years that compactification of type IIA~or~B
strings on Calabi-Yau three-folds has an $N=2$, $D=4$ field theory limit.
(See for example \cite{cand,dewitIIB,seiberg,cmap}
and for explicit constructions,
\cite{shortsab,bodcada,bodcadv,bodcadb}.)  The 10-dimensional 
bosonic field content, in the
electric description of type IIB strings, consists of the
Neveu-Schwarz-Neveu-Schwarz (NS-NS) metric
($\hat{g}_{\hat{\mu} \hat{\nu}}$), dilaton ($\hat{\varphi}$) and 
two-form potential ($\hat{B}^{(1)}_{\hat{\mu} \hat{\nu}}$) and the
Ramond-Ramond (RR) dilaton ($\hat{l}$), two-form potential
($\hat{B}^{(2)}_{\hat{\mu} \hat{\nu}}$) and four-form potential
($\hat{D}_{\hat{\mu} \hat{\nu} \hat{\rho} \hat{\sigma}}$).  (The hats
distinguish the 10-dimensional fields/indices from 4-dimensional ones.)
Under
compactification on a Calabi-Yau three-fold, the metric gives rise to
$2h_{21}+h_{11}$ real scalars and the $D=4$ spacetime metric; each dilaton
gives another scalar; each two-form potential gives rise to $h_{11}+1$
real scalars and the four-form potential gives $h_{11}$ real scalars and
$h_{21}+1$ vectors (and their duals)\cite{bodcada,bodcadv,bodcadb}.
This is the bosonic field content
of $N=2$, $D=4$ supergravity with $h_{21}$ vector multiplets,
$h_{11}+1$ hypermultiplets and a gravity multiplet which contains the
graviphoton which comes from $\hat{D}_{\hat{\mu} \hat{\nu} \hat{\rho}
\hat{\sigma}}$
aligned along the Calabi-Yau holomorphic
three-form\cite{bodcadv,special0,special1,billo}.
Here the $h_{pq}$ are the Hodge numbers of the complex, K\"{a}hler manifold. 

In this paper the consequences of giving expectation values to the field
strengths of the
10-dimensional fields are examined.  From Lorentz
invariance only the three-form field strengths can get expectation
values, since on a generic (i.e. not $T^6$ or $K_3\times T^2$)
Calabi-Yau, $h_{10}$=0.
In section {\ref{expect}} it is shown that, as in \cite{joenandy}, 
giving the field strengths expectation values corresponds, under
dimensional reduction, to giving electric and magnetic charges to the
dilaton hypermultiplet.  In principle, the consistency (under the
10-dimensional equations of motion) of the expectation values with the
Calabi-Yau structure of the compactification should be
examined. However,
since string theory suppresses the interactions of RR fields by a
factor of the string coupling constant $e^{\hat{\varphi}}$, if
attention is restricted to the weak coupling limit, where string
perturbation theory is valid, then the theory for non-zero RR
expectation values is just a
perturbation of the usual Calabi-Yau compactification.
Similarly, the field equation coupling the NS-NS field to gravity is
suppressed by the volume of the Calabi-Yau, so the large Calabi-Yau
volume limit will be taken.

When the dilaton hypermultiplet is charged, the classical potential
of the theory becomes non-trivial~\cite{joenandy,special1,special2,%
special3}.  In \cite{joenandy}, it was shown that giving a
Calabi-Yau-valued expectation
value to the RR 10-form field strength in the IIA theory,
resulted in a potential with
no non-singular minima;
the theory was driven either to conifold points---where including
fields corresponding
to massless black holes removes the singularity~%
\cite{coniandy,coni,greenes}---or to
the infinite Calabi-Yau-volume
limit.  It was subsequently speculated in \cite{kallosh2} that
if an additional RR field strength was given an expectation value,
that the potential would have a non-singular minimum.
It will be shown in section~%
\ref{magminima} that on the IIB side, both RR and NS-NS field strengths
must have expectation values in order for the potential to have a minimum.
This can be understood as follows.  As explained in more detail
in section \ref{expect}, the RR and NS-NS three-form
expectation values are elements of $H^3(CY;\IZ)$, the natural basis for
which is defined up to an
$Sp(h_{21}+1;\IZ)$ transformation (see, for example, \cite{andyspec,bodcadv}).
Thus, the basis can be rotated so that
the RR field strength is aligned along a specific basis vector.
If the NS-NS field strength vanishes, 
this theory is related by mirror symmetry
to that described in \cite{joenandy} for which,
as just stated, the potential has no minimum.  Hence, having only
RR expectation values is insufficient for the potential to have
a minimum.\footnote{The argument can be given without reference to
the IIB theory by considering the $Sp(h_{11}+1;\IZ)$ action on
the anticommuting basis for $H^0 \oplus H^2 \oplus H^4 \oplus H^6$
\cite{canossa}.}

While the potential can have minima when both RR and NS-NS
fields have expectation values, it turns out that the minima not only occur
for values of the moduli that are outside the region of validity
(described above) of the
calculation, but are not supersymmetric and hence are not protected
from quantum corrections.  At a conifold singularity, or at the more general
singularities
(Argyres-Douglas points)
of~\cite{adpts}, the potential can have flat directions with
$N=2$ supersymmetry.  No $N=1$ supersymmetric minima were found in this
paper.  This contrasts with~\cite{susywitten}
where $N=1$, $D=4$ type IIB vacua were found with non-trivial superpotential,
by considering non-perturbative effects in compactifications
on complex manifolds that were not necessarily Calabi-Yau, and where
the coupling constant varied over the Calabi-Yau.

The paper is organized as follows.
In section \ref{expect} it is
shown how electric and magnetic charges for the dilaton hypermultiplet
arise from expectation values of the three-form field strengths.  This
uses some results from \cite{bodcadb} which are reviewed in the
appendix.  These results are then used in 
section \ref{classpotsect} to derive the classical potential for the theory.
The formula for the classical potential in
a general $N=2$ supergravity theory is reviewed in section~%
\ref{reviewandextend}.  The formulas in the literature \cite{special1,%
special2,special3} all hold only in the absence of magnetic charge; to the
author's
knowledge, a magnetic formula does not exist in the literature.  One
is proposed at the end of section \ref{reviewandextend}.  It turns out
that
the pure electric potential contains, for the purposes of this calculation,
many of the same
features as the general one,
but is much simpler.
Therefore the
electric potential is discussed in detail before the magnetic one.
  (Of course, the electric potential cannot be used
for the analysis at Argyres-Douglas points.)
Assumptions of the model (such as the absence of the holomorphic
prepotential for the vector moduli) are discussed in section~%
\ref{elpotsect} and then the electric potential for the model under
consideration is given explicitly.  An explicit expression for
the general potential is then discussed in section~\ref{magpotsect}.
The electric potential is minimized in section~\ref{minima} and the
general potential is minimized in section~\ref{magminima}.
Supersymmetric minima at conifold points and particularly
Argyres-Douglas points are discussed in section~%
\ref{conifold}.  Section \ref{conc} is the conclusion.

\section{Dilaton charges} \label{expect}
Although the self-duality of the five-form field strength in type IIB
string theory implies that the latter cannot be described by a supersymmetric
10-dimensional action, the bosonic fields can be
described by a non-self-dual action in which the equation of
motion for the five-form field strength is replaced by its Bianchi
identity\cite{nsd}.  This is consistent with self-duality, but does not
imply it.  When self-duality is imposed as a
compactification condition, the non-self-dual
action yields the correct compactified
theory\cite{nsd,fields}.
In the Einstein frame, the action is,\cite{nsd}
\begin{eqnarray} \label{nsdlag}
S = \int d^{10}\hat{x} \sqrt{-\hat{g}} &
\left\{ \frac{1}{2} \hat{R}
     -\frac{1}{8} {\rm Tr}(\partial_{\hat{\mu}}\hat{\cal M}
           \partial^{\hat{\mu}}\hat{\cal M}^{-1})
     +\frac{3}{8} \hat{\cal H}^{\rm T}_{\hat{\mu}\hat{\nu}\hat{\rho}}
           \hat{\cal M}
           \hat{\cal H}^{\hat{\mu}\hat{\nu}\hat{\rho}}
     +\frac{5}{12} \hat{F}^2_{\hat{\mu} \hat{\nu} \hat{\rho}
           \hat{\sigma} \hat{\tau}} + \right. \nonumber \\
&  \left. \frac{1}{192} \varepsilon^{ij} \hat{\varepsilon}^{\hat{\mu}
           \hat{\nu}
           \hat{\rho} \hat{\sigma} \hat{\tau} \hat{\lambda} \hat{\alpha}
           \hat{\beta} \hat{\gamma} \hat{\delta}} \hat{D}_{\hat{\mu}
           \hat{\nu} \hat{\rho} \hat{\sigma}} \hat{H}^{(i)}_{\hat{\tau}
           \hat{\lambda} \hat{\alpha}} \hat{H}^{(j)}_{\hat{\beta} \hat{\gamma}
           \hat{\delta}} \right\}.
\end{eqnarray}
The field definitions are
\begin{mathletters} \label{fielddefs}
\begin{eqnarray} 
\label{defM}
\hat{\cal M} = \frac{1}{{\rm Im} \hat{\lambda}}\left( \begin{array}{cc}
     |\hat{\lambda}|^2 & -{\rm Re} \hat{\lambda} \\
     -{\rm Re} \hat{\lambda} &        1 \end{array} \right);&
     \hat{\lambda} = \hat{l} + i e^{-\hat{\varphi}}, \\
\nonumber \\
\label{defH}
\hat{\cal H}_{\hat{\mu}\hat{\nu}\hat{\rho}} =
     \left( \begin{array}{c} \hat{H}^{(1)}_{\hat{\mu}\hat{\nu}\hat{\rho}}
     \\ \hat{H}^{(2)}_{\hat{\mu}\hat{\nu}\hat{\rho}}
     \end{array} \right) ;& {\hat{H}^{(i)}_{\hat{\mu}\hat{\nu}\hat{\rho}}
     =  \partial_{[\hat{\mu}}\hat{B}^{(i)}_{\hat{\nu}\hat{\rho}]}}, \\
\nonumber \\
\label{defF}
{\hat{F}}_{\hat{\mu}\hat{\nu}\hat{\rho}\hat{\sigma}\hat{\tau}} =
     \partial_{[\hat{\mu}}\hat{D}_{\hat{\nu}\hat{\rho}\hat{\sigma}\hat{\tau}]}
     + \frac{3}{4}\varepsilon^{ij} \hat{B}^{(i)}_{[\hat{\mu}\hat{\nu}} 
     \partial_{\hat{\rho}}\hat{B}^{(j)}_{\hat{\sigma}\hat{\tau}]}.
\end{eqnarray}
\end{mathletters}
Also, $\hat{\varepsilon}_{0\ldots 9}=\sqrt{-g}$.
It is in the final term of equation~(\ref{nsdlag}) that the $D=4$
vectors (from $\hat{D}$) interact with the $D=4$ scalars from the
three-form field strengths.%
\footnote{Another contribution, of opposite sign but not equal magnitude,
of this form arises
from the $\hat{F}^2$ term because of self-duality of $\hat{F}$ and
the Chern-Simons term in equation~(\ref{defF}): $\hat{F}^2 \propto
\hat{F}\wedge\hat{F} \propto \varepsilon^{ij}\hat{F}\wedge\hat{H}^{(i)}
\wedge\hat{B}^{(j)}+\ldots$; see equation~(\ref{usef}).  The remaining
terms do not involve $\hat{H}^{(i)}$ and therefore will not yield a new
interaction.}
Therefore it is this term that will be examined closely.

It is convenient to rewrite this (up to an overall constant) as
\begin{equation} \label{usef}
\varepsilon^{ij} \hat{\int} \hat{F} \wedge \hat{H}^{(i)} \wedge \hat{B}^{(j)}.
\end{equation}
(The Chern-Simons terms in equation~(\ref{defF}) don't contribute
because of the (anti)-symmetry of the wedge product.)  To compactify
to four dimensions use \cite{bodcadv,summary,billo}
\begin{equation} \label{foncy}
\hat{F} = F^\Lambda \wedge \alpha_\Lambda - G_\Lambda \wedge \beta^\Lambda
      + \ldots,
\end{equation}
where $(\alpha_\Lambda, \beta^\Lambda), \Lambda = \{0,1,\ldots,h_{21}\}$
are some choice of symplectic basis for
$H^3({\rm CY})$, $F^\Lambda_{\mu \nu}$ are the 4-dimensional vector field
strengths and $G_{\Lambda \mu \nu}$ are the magnetic field strengths
\cite{bodcadv,summary,billo} and the dual relationship between
$F^\Lambda_{\mu \nu}$ and
$G_{\Lambda \mu \nu}$ is due to self-duality of $\hat{F}$;
the terms that have been left out of
equation (\ref{foncy}) are those which will not contribute to the integral
in equation (\ref{usef}).  The 3-form field strengths
are given Calabi-Yau expectation values via
\begin{mathletters} \label{hexpect}
\begin{eqnarray} \label{hexpect1}
<\hat{H}^{(1)}> & = & \nu_{e\Lambda}^{(1)} \beta^\Lambda - 
        \nu_m^{(1)\Lambda} \alpha_\Lambda, \\ \label{hexpect2}
<\hat{H}^{(2)}> & = & \nu_{e\Lambda}^{(2)} \beta^\Lambda - 
        \nu_m^{(2)\Lambda} \alpha_\Lambda,
\end{eqnarray}
\end{mathletters}
where the $\nu_{m(e)}$ are constants that have been prematurely
identified as values of the magnetic (electric) charges.  

Using equations (\ref{hexpect}),
integration of equation~(\ref{usef}) over the Calabi-Yau gives
\begin{equation} \label{substitute}
\varepsilon^{ij} \int \left({\nu}^{(i)}_{e\Lambda}F^{\Lambda}\wedge
      B^{(j)} - {\nu}^{(i)\Lambda}_m G_{\Lambda}\wedge B^{(j)} \right).
\end{equation}
Writing $F^{\Lambda}$ and $G_{\Lambda}$ in terms of electric and
magnetic vector potentials $A^{\Lambda}_\mu$ and $\tilde{A}_{\Lambda\mu}$,
gives, after an integration by parts, (again up to a constant)
\begin{equation} \label{dual}
\varepsilon^{ij} \int d^4x \sqrt{-g}
     \left({\nu}^{(i)}_{e\Lambda}A_\mu^\Lambda H^{(j)\mu} -
     {\nu}^{(i)\Lambda}_m\tilde{A}_{\Lambda\mu}H^{(j)\mu} \right),
\end{equation}
where
\begin{equation} \label{defhdual}
H^{(i)\mu} = \frac{1}{6} \varepsilon^{\nu\sigma\tau\mu}\partial_\nu
     B^{(i)}_{\sigma\tau}.
\end{equation}

To understand this, it is necessary to relate the $H^{(i)}_\mu$ to the
4-dimensional scalars.  This is done in the appendix.  The result is
that to lowest order in the coupling constant, with, for simplicity, the
fields corresponding to the $h_{21}$ data set to zero,
\begin{mathletters} \label{whatishdual}
\begin{eqnarray} 
\label{whatishdual1}
H^{(1)}_\mu & = \frac{2}{3} e^{\frac{5}{4}\tilde{K}
      -\frac{1}{4}K} \partial_\mu \mbox{Im} S, \\
\label{whatishdual2}
H^{(2)}_\mu & = \frac{2 \sqrt{2}}{3} e^{\frac{\tilde{K}}{4}+\frac{3K}{4}}
      \partial_\mu
      \mbox{Im} C_0;
\end{eqnarray}
also, the string coupling constant is
\begin{equation} \label{whatisexpphi}
e^{\hat{\varphi}} = \sqrt{2} e^{\frac{\tilde{K}}{2}-\frac{K}{2}}.
\end{equation}
\end{mathletters}
Here $S$ and $C_0$ are the
$N=1$ superfields 
which form the dilaton hypermultiplet; as in \cite{joenandy} the four
dimensional dilaton has been generalized to
\begin{equation} \label{gen4dil}
\phi = \frac{1}{2} e^{-\tilde{K}}.
\end{equation}
The K\"{a}hler potential of the
special geometry precurser to the quaternionic manifold
\cite{cmap,fersab} is denoted by $K$, while the K\"{a}hler potential for
the rest of the quaternionic manifold is denoted by $\tilde{K}$.  (In
particular, the metric on the hypermultiplets is determined by
$\tilde{K}$.)  
In equation~(\ref{gen4dil}), the non-dilaton multiplets are omitted from
the K\"{a}hler  potential (see also equation~(\ref{whatK})).
That is, the above equations were derived by
explicitly compactifying on a ``minimal'' Calabi-Yau manifold with
$h_{11}$=1 and $h_{21}$=0 and even ignoring much of this Calabi-Yau
data.  For this compactification there is a relation between $\mbox{Im}
Z$ ($Z$
being the complex coordinate on the one complex-dimensional special
geometry precurser to the quaternionic manifold), $K$ and
${\cal R}^{00}$ (where
${\cal R}^{00}$ is defined in the appendix).  Since for more generic manifolds,
there are many $\mbox{Im} Z$s while there is only one $K$ (or ${\cal R}^{00}$),
it is preferable to use the latter in these formulas.

Substituting equations (\ref{whatishdual})
into equation~(\ref{dual}) gives, after a Weyl
rescaling $g_{\mu\nu} \rightarrow \sqrt{2} e^{\frac{3\tilde{K}}{4}
+\frac{K}{4}}
g_{\mu\nu}$ (to go to the $D=4$ Einstein metric),
\begin{eqnarray} \label{interaction}
\frac{2 \sqrt{2}}{3} \int d^4x \sqrt{-g} &
     \left\{\sqrt{2}
          e^{\tilde{K}+K}\nu^{(1)}_{e\Lambda}A_\mu^\Lambda
          \partial^\mu {\rm Im} C_0 -
          \sqrt{2}
          e^{\tilde{K}+K}\nu^{(1)\Lambda}_m\tilde{A}_{\mu\Lambda}
          \partial^\mu {\rm Im} C_0 - \right. \nonumber \\
& \left. e^{2 \tilde{K}}\nu^{(2)}_{e\Lambda}A^\Lambda_\mu
          \partial^\mu \mbox{Im} S +
          e^{2 \tilde{K}}\nu^{(2)\Lambda}_m\tilde{A}_{\Lambda\mu}
          \partial^\mu \mbox{Im} S\right\}.
\end{eqnarray}
This can be recognized as the interaction terms of the vector potentials
with charged fields $e^S$ and $e^{C_0}$.  Hence, as in \cite{joenandy},
completing the square with the kinetic terms for the hypermultiplets
\cite{fersab,bodcadb} gives (with an
appropriate numerical rescaling of $\nu^{(i)}_{e(m)\Lambda}$)
\begin{eqnarray} \label{finalterm}
S = \int d^4x \sqrt{-g} &\left\{ 8 e^{\tilde{K}+K} \left(
        \nu_{e\Lambda}^{(1)}A_{\mu}^{\Lambda}-
        \nu_m^{(1)\Lambda}\tilde{A}_{\Lambda\mu} +
        \partial_\mu {\rm Im} C_0 \right)^2 \right. \nonumber \\
&\left. + 2 e^{2\tilde{K}} \left(\nu_{e\Lambda}^{(2)}A_{\mu}^{\Lambda}-
        \nu_m^{(2)\Lambda}\tilde{A}_{\Lambda\mu} +
        \partial_\mu \mbox{Im} S \right)^2 + \ldots \right\}.
\end{eqnarray}
From this equation, it is seen that ${\rm Im}C_0$ carries electric
(magnetic) charges $\nu^{(1)}_{e\Lambda}$ ($\nu^{(1)\Lambda}_m$) and
that ${\rm Im}S$ carries electric (magnetic)
charges $\nu^{(2)}_{e\Lambda}$ ($\nu^{(2)\Lambda}_m$).  Note that this
coincides with the type IIA calculation of
\cite{joenandy} where ${\rm Im}S$ carried
electric and magnetic charges proportional to the expectation values of
the RR field strengths of the IIA theory (the authors of \cite{joenandy}
did not consider NS-NS
expectation values).

Charge quantization \cite{dbranes,joenandy}
requires that $\nu^{(i)}_{e\Lambda}$ and
$\nu^{(i)\Lambda}_m$ be integers.%
\footnote{Presumably this result can also
be obtained directly from equations (\ref{hexpect})
and quantization of RR charge in ten dimensions \cite{nep,teit,dbranes}
(S duality extends this quantization to
the NS-NS three-form charge).}

Finally, note that an $Sp(h_{21}+1,\IZ)$ transformation on the basis
$(\beta^\Lambda,\alpha_\Lambda)$ will rotate the
charge vectors $(\nu^{(i)\Lambda}_m,\nu^{(i)}_{e\Lambda})$.  
By performing $SL(2,\IZ)$ electromagnetic duality transformations on each
vector independently,
followed by a perturbative
$Sp(h_{21}+1,\IZ)$ transformation of the form $\left(
\begin{array}{cc} \bbox{\sf A}_{h_{21}+1} & 0 \\ 0 &
\bbox{\sf (A}_{h_{21}+1}^{\bbox{\sf T}}\bbox{\sf )^{-1}} \end{array}
\right), \bbox{\sf A}_{h_{21}+1} \in SL(h_{21}+1,\IZ)$
it is always possible to perform a rotation so that the NS-NS charge
vector $(\nu^{(1)\Lambda}_m,\nu^{(1)}_{e\Lambda})$ is pure electric and
aligned with respect to only one $U(1)$, say $A^0$, with
positive charge.
It is then possible to perform further electro-magnetic duality transformations
on all the vectors but $A^0$ so that the only potentially non-zero magnetic
charge is $\nu^{(2)0}_m$.  This can be followed by a perturbative
$Sp(h_{21}+1,\IZ)$ transformation with $\bbox{\sf A}_{h_{21}+1}
= \left( \begin{array}{cc}
1 & 0 \\ 0 & \bbox{\sf A}_{h_{21}} \end{array} \right)$ so that all
$\nu^{(2)}_{e\Lambda}$ vanish except possibly
$\nu^{(2)}_{e0}$ and $\nu^{(2)}_{e1}$.  Furthermore,
if $\nu^{(1)}_{e\Lambda} \nu^{(2)\Lambda}_m - \nu^{(1)\Lambda}_m
\nu^{(2)}_{e \Lambda} = 0$ then in the new basis
$\nu^{(2)0}_m = 0$ and all charges are
pure electric.  This is known~\cite{adpts} as the local case.  Otherwise,
$\nu^{(2)0}_m \neq 0$; this is the non-local case.

To summarize the last paragraph,
it is always possible to choose a symplectic basis so that
the NS-NS charge vector $(\nu^{(1)\Lambda}_m, \nu^{(1)}_{e\Lambda})$
is pure electric, with respect to only one $U(1)$ and the RR charge vector
is, at most, magnetically and electrically charged with respect to that
$U(1)$ and electrically charged with respect to one other $U(1)$.
In fact, there is more freedom in special cases.
In the local (vanishing magnetic charge)
case, if $\nu^{(2)}_{e0} = m \nu^{(2)}_{e1}, m \in \IZ$, then choosing
$\bbox{\sf A}_{h_{21}+1}
= \left(\begin{array}{ccc} 1 & -m & 0 \\ 0 & 1 & 0 \\
0 & 0 & \I1 \end{array} \right)$ makes the RR charge vector electrically
charged under only one $U(1)$, different from that under which the NS-NS
is charged.  (This does not work in the non-local case as the magnetic
charge transforms non-trivially.)  In the non-local case if
$\nu^{(2)}_{e\Lambda}$ are integer multiples of $\nu^{(2)0}_m$, then
they can be eliminated using the symplectic matrix (all but
$\Lambda=0,1$ components are suppressed) $\left (
\begin{array}{cccc} 1 & 0 & -\frac{\nu^{(2)}_{e0}}{\nu^{(2)0}_m} &
-\frac{\nu^{(2)}_{e1}}{\nu^{(2)0}_m} \\
0 & 1 & -\frac{\nu^{(2)}_{e1}}{\nu^{(2)0}_m} & \hbox{arbitrary} \\
0 & 0 & 1 & 0 \\ 0 & 0 & 0 & 1 \end{array} \right)$ so that the RR charge
vector is pure magnetic under the same $U(1)$ that the NS-NS charge
vector is pure electric.

\section{Classical Potential} \label{classpotsect}
\subsection{Review of $N=2$ Supergravity with Electrically Charged Matter
and Generalization to Magnetically Charged Matter} \label{reviewandextend}
Recall \cite{orign2} that
there are essentially three types of $N=2$ multiplets: gravitational, vector
and hypermultiplets.  The gravitational multiplet consists of the
graviton $g_{\mu \nu}$, two gravitini $\psi^A_\mu$ and the graviphoton $A_\mu$.
The gravitini form a doublet under the $SU(2)$ which
relates the two supersymmetries; hence they are labelled by the index
$A=1,2$.  Each of the $h_{21}$ vector multiplets
consists of a vector $A^a_\mu$, two gauginos $\lambda^{aA}$ and a complex
scalar $z^a$, $a=1\ldots h_{21}$.  (Vector multiplets can also be written
in $N=1$ superfield notation as a chiral multiplet plus an $N=1$ vector
multiplet.) The $h_{11}+1$ hypermultiplets consist of
$4(h_{11}+1)$ scalars $q^u$ and $2h_{11}$ hyperini $\zeta_\alpha$, $u=1\ldots
4(h_{11}+1)$ and $\alpha=1\ldots2(h_{11}+1)$.
The natural way in which the index
$\alpha$ arises will be discussed shortly.  (As used in the previous
section, in $N=1$ superfield notation, a hypermultiplet consists
of two chiral superfields.)

The vector multiplet scalars map out a special K\"{a}hler
manifold.  This will only be described briefly here; for
more detail the reader is referred to the literature~\cite{andyspec,%
special0,special1,special2,special3,noprepot,vanderall,summary}.
Roughly, a special K\"{a}hler manifold is a complex K\"{a}hler manifold whose
K\"{a}hler potential is derived from a holomorphic prepotential $F(z^a)$.
It is convenient to define special coordinates via projective coordinates
$X^\Lambda, \Lambda=0\ldots h_{21}, z^a=\frac{X^a}{X^0}$.  Then,
$F(X^\Lambda)$ is required to be a homogeneous function of degree 2.
Defining
\begin{equation} \label{defprepot}
F_\Lambda = \frac{\partial F(X^\Lambda)}{\partial X^\Lambda},
\end{equation}
the K\"{a}hler potential can be written as
\begin{equation} \label{kv}
K_V = -\ln i(\bar{X}^\Lambda F_\Lambda - X^\Lambda \bar{F}_\Lambda).
\end{equation}
Equation~(\ref{kv}) is $Sp(h_{21}+1,\IR)$ invariant, where $(X^\Lambda,
F_\Lambda)$ transform as a symplectic vector.  In special coordinates
it is natural, especially given the superspace Bianchi identities
\cite{special0}, to define the graviphoton to be $A^0$, so that, in
addition to the $X^\Lambda$s there are $A^\Lambda$s.  (In fact, this
argument in reverse is the usual reason for introducing $X^0$.)
Thus, it is seen how the symplectic formulation of special K\"{a}hler
geometry is the natural one.  In fact, the $Sp(h_{21}+1,\IR)$
transformation is the same one that mixes the basis vectors $\alpha_\Lambda,
\beta^\Lambda$ of $H^3(CY)$.  Charge quantization, and/or the requirement
that ($\beta^\Lambda, \alpha_\Lambda$) be in $H^3(CY,\IZ)$, requires the
restriction to $Sp(h_{21}+1,\IZ)$.  
It follows immediately that in a general basis
$A^0$ will be the graviphoton only if the Calabi-Yau holomorphic three-form
is aligned with $\alpha_0$; in general this is not only false but impossible.
Also,
the $X^\Lambda(z^a)$ are not necessarily projective versions of
the coordinates $z^a$, but are general holomorphic functions.
This fact and the fact that the holomorphic prepotential $F(X^\Lambda)$
is not guaranteed to exist in a general basis, makes it necessary to
find symplectic invariant, prepotential independent, formulas for quantities.
This has been done in \cite{noprepot,summary}.

It is sometimes useful to define
\begin{equation} \label{defLandM}
L^\Lambda = e^{\frac{K_V}{2}}X^\Lambda;
M_\Lambda = e^{\frac{K_V}{2}}F_\Lambda.
\end{equation}
The natural derivative to use for these is the covariant derivative
\begin{mathletters} \label{defcovderivLM}
\begin{eqnarray} 
\label{deffLambdaa}
\nabla_a L^\Lambda &= \partial_a L^\Lambda +
    \frac{1}{2} (\partial_a K_V) L^\Lambda &\equiv f^\Lambda_a; \partial_a = 
    \frac{\partial}{\partial z^a}, \\
\label{defhLambdaa}
\nabla_a M_\Lambda &= \partial_a M_\Lambda +
    \frac{1}{2} (\partial_a K_V) M_\Lambda &\equiv h_{a\Lambda}, \\
\label{deffLambdaabaris0}
\nabla_{\bar{a}} L^\Lambda &= \partial_{\bar{a}} L^\Lambda -
    \frac{1}{2} (\partial_{\bar{a}} K_V) L^\Lambda &\equiv 0, \\
\label{defhLambdaabaris0}
\nabla_{\bar{a}} M_\Lambda &= \partial_{\bar{a}} M_\Lambda -
    \frac{1}{2} (\partial_{\bar{a}} K_V) M_\Lambda &\equiv 0,
\end{eqnarray}
\end{mathletters}
where the last two equations follow from holomorphicity of $X^\Lambda$ and
of $F_\Lambda$.  Supersymmetry implies the existence of a matrix 
${\cal N}_{\Lambda \Sigma}$
so that \cite{vanderall,noprepot,summary}
\begin{mathletters} \label{defN}
\begin{equation} \label{defNforM} 
M_\Lambda = {\cal N}_{\Lambda \Sigma} L^\Sigma,
\end{equation}
and
\begin{equation} \label{defNforh}
h_{a\Lambda} = \bar{\cal N}_{\Lambda \Sigma} f_a^\Sigma.
\end{equation}
\end{mathletters}
Finally, it is convenient to define
\begin{equation} \label{defU}
U^{\Lambda \Sigma} = g^{a\bar{b}} f_a^\Lambda \bar{f}_{\bar{b}}^\Sigma,
\end{equation}
where $g^{a \bar{b}}$ is the inverse of the K\"{a}hler metric $g_{a \bar{b}}
= \partial_a \partial_{\bar{b}} K_V$.

The hypermultiplets parametrize a quaternionic manifold.  A quaternionic
manifold has three almost complex structures that obey the quaternionic
($Sp(1)\sim SU(2)$)
algebra and whose K\"{a}hler forms are covariantly closed
using an $SU(2)$ connection whose field strength is proportional to
the K\"{a}hler form triplet.
That is,
\begin{mathletters} \label{defomegas}
\begin{eqnarray} 
\label{l2}
&\Omega^{xu}{}_{v}\Omega^{yv}{}_{w} = -\delta^{xy} \delta^u{}_v -
     \varepsilon^{xyz}\Omega^{zu}{}_{w}, \\
\label{dOmega}
&\nabla \Omega^x \equiv
d\Omega^x + \varepsilon^{xyz} \omega^y \wedge \Omega^z = 0,&{\rm and} \\
\label{domega}
&d\omega^x + \frac{1}{2}\varepsilon^{xyz} \omega^y \wedge \omega^z = \Omega^x,
\end{eqnarray}
\end{mathletters}
using the canonical normalization%
\cite{bagwit,special1,special2,special3},
where $\Omega^{xu}{}_{v}$ is the triplet of complex structures and
$\omega^x_u$ is the $SU(2)$ connection, $x=1,2,3$.
The holonomy of a quaternionic manifold is $SU(2)\times H$ with $H \subset
Sp(h_{11}+1)$.  The $SU(2)$ factor is that whose curvature is the K\"{a}hler
form triplet and is also the $SU(2)$ that rotates the supersymmetries.
From the holonomy of the manifold, the natural flat metric is the $SU(2)\times
Sp(h_{11}+1)$ one;
i.e.\ the vielbein is $U^u_{A\alpha}$ where again $A=1,2$ is the
$SU(2)$ index and $\alpha=1,\ldots,2(h_{11}+1)$ is the $Sp(h_{11}+1)$ index.
Because each hyperino is an $SU(2)$ singlet, the hyperini are labelled
only by the $\alpha$ index,
as indicated above.

If the hypermultiplet is electrically
charged, then there must be a symmetry of the
theory that is gauged.  In other words the vector multiplets
gauge isometries of the quaternionic manifold.  (This is also true of
the special K\"{a}hler manifolds; however, the vectors considered here
are abelian and hence uncharged.)  The covariant derivative of the
coordinate (hypermultiplet scalar) is (compare to equation~(\ref{finalterm}))
\begin{equation} \label{covderivcoord}
\nabla_\mu q^u = \partial_\mu q^u + k^u_\Lambda A^\Lambda_\mu
\end{equation}
where $k^u_\Lambda$ is the Killing vector that generates the isometry.
The isometries of the quaternionic
manifold must respect the quaternionic nature of the manifold.  So,
the Lie derivatives of the K\"{a}hler forms and the $SU(2)$ connection,
with respect to the Killing vector,
must vanish up to an $SU(2)$ gauge transformation \cite{special1,special2,%
special3}.  Then, Killing prepotentials, ${\cal P}^x_\Lambda$, can be found
which satisfy \cite{galicki,special1,special2,special3}
\begin{equation} \label{prepot}
\Omega^x(k_\Lambda,\cdot) = -d{\cal P}^x_\Lambda - \varepsilon^{xyz}
     \omega^y {\cal P}^z_\Lambda.
\end{equation}

The general formula for the classical potential in an $N=2$ supergravity
theory was given in \cite{wardid1,wardid2}.  Note that the derivation
therein is very general and should hold in the presence of both
electric and magnetic charges.  The potential is given by the Ward identity
\begin{equation} \label{wardeq}
V \delta^A{}_B = g_{a \bar{b}}W^{aAC}\bar{W}^{\bar{b}}_{BC} +
     N^{\alpha A}\bar{N}_{\alpha B} - 12 S_{BC}\bar{S}^{CA},
\end{equation}
where $W^{aAC}$,
$N_\alpha{}^A$ and $S_{AB}$ are respectively the matrices governing the SUSY
transformations of the gaugino, the hyperino and the gravitino mass matrix.
Specifically,
\begin{mathletters} \label{susytransf}
\begin{eqnarray} \label{susygaugino}
&\delta\lambda^{aA} = \ldots + W^{aAB}\epsilon_B, \\
\label{susyhyper}
&\delta\zeta_\alpha = \ldots + N_\alpha^A \epsilon_A, \\
\label{susygrav}
&\delta \psi_{A\mu} = {\cal D}_\mu \epsilon_A +
     \ldots + i S_{AB} \gamma_\mu \epsilon^B,
\end{eqnarray}
\end{mathletters}
where $\epsilon$ is the SUSY transformation parameter,
${\cal D}_\mu$ is
the spacetime covariant derivative, and the missing terms are those
which vanish in the Lorentz invariant, bosonic background.

These matrices were worked out for the case of vanishing magnetic charge
in \cite{special1,special2,special3}.   They are given by
\begin{mathletters} \label{susymatrices}
\begin{eqnarray} \label{matrixW}
&W^{aAB} = i (\sigma_x)_C{}^B \varepsilon^{CA}{\cal P}^x_\Lambda
     g^{a \bar{b}} \bar{f}^\Lambda_{\bar{b}}, \\
\label{matrixN}
&N_\alpha^A = 2 U^A_{\alpha u}k_\Lambda^u \bar{L}^\Lambda, \\
\label{mass}
&S_{AB} = \frac{1}{2} (\sigma_x)_A{}^C\varepsilon_{BC}{\cal
P}_\Lambda^x L^\Lambda.
\end{eqnarray}
\end{mathletters}
This gives 
\begin{equation} \label{pot}
V = 2h_{uv}k^u_\Lambda k^v_\Sigma L^\Lambda \bar{L}^\Sigma
     + (U^{\Lambda \Sigma} - 3 L^\Lambda \bar{L}^\Sigma)
     {\cal P}_\Lambda^x {\cal P}_\Sigma^x
\end{equation}
for the potential, upon insertion into equation~(\ref{wardeq}).  The
quaternionic metric is denoted by $h_{uv}$.

To generalize this to the case of non-vanishing magnetic charge, it is
necessary (though not necessarily sufficient) to find symplectic invariant
versions of e.g. equations~(\ref{susymatrices}) and (\ref{pot}),
that reduce to these when the magnetic
charge vanishes.  To attempt this, note
first that equation~(\ref{finalterm}) suggests that equation~%
(\ref{covderivcoord}) be replaced by
\begin{equation} \label{magcovderivcoord}
\nabla_\mu q^u = \partial_\mu q^u + k^u_\Lambda A^\Lambda_\mu -
\tilde{k}^{\Lambda u} \tilde{A}_{\Lambda \mu},
\end{equation}
where $\tilde{k}^{\Lambda u}$ is the Killing vector gauged by the magnetic
vectors, $\tilde{A}_\Lambda$, and the minus sign comes from the symplectic
metric.  (Of course this only works on-shell---the off-shell Lagrangian is
necessarily either non-local or non-Lorentz covariant---see e.g.~%
\cite{schwarzsen,new}.)
It is clear, then, (or at least natural to assume) that
$(\tilde{k}^\Lambda, k_\Lambda)$ and $(A^\Lambda, \tilde{A}_\Lambda)$
are symplectic vectors.  It is also clear that the magnetic Killing vectors
suffer from the same restrictions as the electric ones, and hence the
analogue of equation~(\ref{prepot}) holds for $\tilde{\cal P}^{\Lambda x}$.
So the generalization of equations (\ref{susymatrices}) is
\begin{mathletters} \label{magsusymatrices}
\begin{eqnarray} \label{magmatrixW}
&W^{aAB} = i (\sigma_x)_C{}^B \varepsilon^{CA}
     g^{a \bar{b}} ({\cal P}^x_\Lambda\bar{f}^\Lambda_{\bar{b}} - 
     \tilde{\cal P}^{\Lambda x}\bar{h}_{\bar{b}\Lambda}), \\
\label{magmatrixN}
&N_\alpha^A = 2 U^A_{\alpha u}(k_\Lambda^u \bar{L}^\Lambda -
     \tilde{k}^{\Lambda u}\bar{M}_\Lambda), \\
\label{magmass}
&S_{AB} = \frac{1}{2} (\sigma_x)_A{}^C\varepsilon_{BC}({\cal
P}_\Lambda^x L^\Lambda - \tilde{\cal P}^{\Lambda x} M_\Lambda).
\end{eqnarray}
\end{mathletters}
Inserting this into equation~(\ref{wardeq}) gives
\begin{eqnarray} \label{magpot}
V &=& g^{a \bar{b}} ({\cal P}^x_\Lambda f^\Lambda_a -
    \tilde{\cal P}^{\Lambda x} h_{a \Lambda})
    ({\cal P}_{\Sigma}^x \bar{f}_{\bar{b}}^\Sigma -
     \tilde{\cal P}^{\Sigma x} \bar{h}_{\bar{b} \Sigma}) - 
   3 ({\cal P}^x_\Lambda L^\Lambda -
    \tilde{\cal P}^{\Lambda x} M_\Lambda)
    ({\cal P}_{\Sigma}^x \bar{L}^\Sigma -
     \tilde{\cal P}^{\Sigma x} \bar{M}_\Sigma) + \nonumber \\
&& 2 h_{uv} (k^u_\Lambda L^\Lambda -
    \tilde{k}^{\Lambda u} M_\Lambda)
    (k_{\Sigma}^u \bar{L}^\Sigma -
     \tilde{k}^{\Sigma u} \bar{M}_\Sigma).
\end{eqnarray}
Deriving these using the approach of \cite{special0,special1} would be
the ultimate justification of these formulas.

\subsection{The Local Case} \label{elpotsect}
Returning to the theory at hand,
if $\nu^{(1)}_{e\Lambda} \nu^{(2)\Lambda}_m - \nu^{(1)\Lambda}_m
\nu^{(2)}_{e\Lambda} = 0$ then, as discussed at the end of section~%
\ref{expect}, it is possible to perform an
$Sp(h_{21}+1;\IZ)$ transformation to a basis in which the magnetic charges
vanish.  The more general case of nonvanishing magnetic charge will be
considered in the next subsection.  

From \cite{fersab,joenandy,bodcadb}, the quaternionic structure is given
by
\begin{eqnarray}
\label{whatOmega}
&\Omega^x = i e^\dagger \sigma^x e, & {\rm and} \\
\nonumber \\
\label{whatomega}
&\omega^x \sigma^x = 2i \left( \begin{array}{cc}
     \frac{1}{4}(v-\bar{v}) & -u \\ 
     \bar{u} & -\frac{1}{4}(v-\bar{v}) \end{array} \right),
\end{eqnarray}
where
\begin{mathletters} \label{whatuv}
\begin{eqnarray} \label{whate}
&e = \left(\begin{array}{c}u \\ v \end{array} \right);  \\
\nonumber \\ \label{whatu}
&u = 2 e^{\frac{\tilde{K}}{2}+\frac{K}{2}} dC_0,& {\rm and} \\
\label{whatv}
&v = e^{\tilde{K}}dS - 4 (C_0+\bar{C_0})e^{\tilde{K}+K}dC_0,
     &{\rm with}  \\ \label{whatK}
&\tilde{K} = -\ln  [S+\bar{S}-2 (C_0+\bar{C}_0)^2 e^K ],
\end{eqnarray}
\end{mathletters}
ignoring all but the dilaton multiplet.
This gives,
\begin{mathletters} \label{Omegas}
\begin{eqnarray}
\label{Omega1}
\Omega^1 &=& i (\bar{u} \wedge v + \bar{v} \wedge u) \nonumber \\
&         =& 2 i e^{\frac{3\tilde{K}}{2}+\frac{K}{2}}
            (d\bar{C}_0 \wedge dS + d\bar{S} \wedge dC_0)
            - 16 i (C_0+\bar{C}_0) e^{\frac{3\tilde{K}}{2}+\frac{3K}{2}}
            d\bar{C}_0 \wedge dC_0, \\
\label{Omega2}
\Omega^2 &=& (\bar{u} \wedge v - \bar{v} \wedge u) \nonumber \\
&         =& 2 e^{\frac{3\tilde{K}}{2}+\frac{K}{2}}
            (d\bar{C}_0 \wedge dS - d\bar{S} \wedge dC_0), \\
\label{Omega3}
\Omega^3 &=& i (\bar{u} \wedge u - \bar{v} \wedge v) \nonumber \\
&         =& 4 i e^{\tilde{K}+K}(1 - 4 (C_0 + \bar{C}_0)^2
            e^{\tilde{K}+K}) d\bar{C}_0 \wedge dC_0 -
            i e^{2 \tilde{K}} d\bar{S} \wedge dS + \nonumber \\
&&          4 i (C_0 + \bar{C}_0) e^{2 \tilde{K} + K}
            (d\bar{C}_0 \wedge dS + d\bar{S} \wedge dC_0), 
\end{eqnarray}

\begin{eqnarray} 
\label{omega1}
&\omega^1 = i(\bar{u}-u)
         = 2 i e^{\frac{\tilde{K}}{2}+\frac{K}{2}} (d\bar{C}_0 - dC_0), \\
\label{omega2}
&\omega^2 = u+\bar{u}
         = 2 e^{\frac{\tilde{K}}{2} + \frac{K}{2}} (dC_0 + d\bar{C}_0), \\
\label{omega3}
&\omega^3 = \frac{i}{2}(v-\bar{v})
         = \frac{i}{2} e^{\tilde{K}} (dS - d\bar{S}) - 2 i (C_0 + \bar{C}_0)
             e^{\tilde{K}+K} (dC_0 - d\bar{C}_0). 
\end{eqnarray} 
\end{mathletters}
It is readily verified that equations (\ref{Omegas}) satisfy equations
(\ref{defomegas}), where the quaternionic metric is derived from
equation~(\ref{whatK}) (see equations (\ref{whathuv}) below).

The Killing vectors can be read off of equation~(\ref{finalterm}).  They
are
\begin{equation} \label{killingvect}
k^u_\Lambda = i {\nu}^{(1)}_{e\Lambda}
     \left(\frac{\partial}{\partial C_0} - \frac{\partial}{\partial
           \bar{C}_0}\right)^u +
     i \nu^{(2)}_{e\Lambda} \left(\frac{\partial}{\partial S}
           - \frac{\partial}{\partial \bar{S}} \right)^u.
\end{equation}
It is readily verified that
\begin{mathletters} \label{liederivs}
\begin{equation} \label{lieomega}
\mbox{\pounds}_{k_\Lambda} \omega^x = 0
\end{equation}
which implies
\begin{equation} \label{lieOmega}
\mbox{\pounds}_{k_\Lambda} \Omega^x = 0,
\end{equation}
\end{mathletters}
where $\mbox{\pounds}_X$ denotes the Lie derivative with respect to the
vector field $X$.
These together imply the vanishing of the $SU(2)$ compensator 
associated with the $k_\Lambda$,
which, in turn, gives~\cite{special1}
\begin{equation} \label{easyP}
{\cal P}^x_\Lambda  = \omega^x_u k^u_\Lambda,
\end{equation}
or
\begin{mathletters} \label{ps}
\begin{eqnarray}
\label{p1}
&{\cal P}^1_\Lambda = 4 e^{\frac{\tilde{K}}{2}+\frac{K}{2}}
        {\nu}^{(1)}_{e\Lambda}, \\
\label{p2}
&{\cal P}^2_\Lambda = 0,& {\rm and} \\
\label{p3}
&{\cal P}^3_\Lambda = 4 (C_0 + \bar{C}_0) e^{\tilde{K}+K}
     {\nu}^{(1)}_{e \Lambda} -  e^{\tilde{K}} \nu^{(2)}_{e\Lambda}.
\end{eqnarray}
\end{mathletters}
These, of course, satisfy equation~(\ref{prepot}).
Again, this coincides with the IIA result of \cite{joenandy} when
$\nu^{(1)}_{e\Lambda}=0$.  

From equation~(\ref{whatK}) the quaternionic metric components are found
to be
\begin{mathletters} \label{whathuv}
\begin{eqnarray} \label{whathss}
h_{S\bar{S}} & = & e^{2 \tilde{K}}, \\ \label{whathsc} \label{whathcs}
h_{S\bar{C}_0} & = & h_{C_0\bar{S}} = -4 (C_0 + \bar{C}_0) e^{2\tilde{K}+K},
\\ \label{whathcc}
h_{C_0\bar{C}_0} &=& 4 e^{\tilde{K}+K} + 16 (C_0 + \bar{C}_0)^2
     e^{2 \tilde{K} + 2K},
\end{eqnarray}
\end{mathletters}
and so
\begin{eqnarray} \label{hkk}
h_{uv}k_\Lambda^u k_\Sigma^v & = & 
      8 e^{\tilde{K}+K}[1+4(C_0+\bar{C}_0)^2e^{\tilde{K}+K}]
      \nu_{e\Lambda}^{(1)} \nu_{e\Sigma}^{(1)} + 2 e^{2 \tilde{K}}
      \nu_{e\Lambda}^{(2)} \nu_{e\Sigma}^{(2)} - \nonumber \\
&&    8 (C_0+\bar{C}_0)
      e^{2\tilde{K} + K}
      (\nu_{e\Lambda}^{(1)} \nu_{e\Sigma}^{(2)} + \nu_{e\Lambda}^{(2)}
       \nu_{e\Sigma}^{(1)}).
\end{eqnarray}
However, from equations (\ref{ps}), it is found that
\begin{eqnarray} \label{pxpx}
{\cal P}^x_\Lambda {\cal P}^x_\Sigma &=& 16 e^{\tilde{K}+K}[1+
           (C_0+\bar{C}_0)^2e^{\tilde{K}+K}]
      \nu_{e\Lambda}^{(1)} \nu_{e\Sigma}^{(1)} + 
      e^{2 \tilde{K}}
      \nu_{e\Lambda}^{(2)} \nu_{e\Sigma}^{(2)} - \nonumber \\
&&    4 (C_0+\bar{C}_0)
      e^{2\tilde{K} + K}
      (\nu_{e\Lambda}^{(1)} \nu_{e\Sigma}^{(2)} + \nu_{e\Lambda}^{(2)}
       \nu_{e\Sigma}^{(1)}).
\end{eqnarray}

Inserting equations
(\ref{hkk}) and (\ref{pxpx}) into equation~(\ref{pot}) gives the classical
potential: 
\begin{eqnarray} \label{whatpot}
V &=& \left\{ 16 e^{\tilde{K}+K}
      [1+4(C_0+\bar{C}_0)^2 e^{\tilde{K}+K}]
      \nu_{e\Lambda}^{(1)} \nu_{e\Sigma}^{(1)} + 4 e^{2 \tilde{K}}
      \nu_{e\Lambda}^{(2)} \nu_{e\Sigma}^{(2)} - \right. \nonumber \\
&&    \left. 16 e^{2\tilde{K}+K} (C_0+\bar{C}_0)
      (\nu_{e\Lambda}^{(1)} \nu_{e\Sigma}^{(2)} + \nu_{e\Lambda}^{(2)}
       \nu_{e\Sigma}^{(1)})\right\}L^\Lambda \bar{L}^\Sigma + \nonumber \\
&&     (U^{\Lambda \Sigma} - 3 L^\Lambda \bar{L}^\Sigma) \left\{
      16 e^{\tilde{K}+K}
      [1+(C_0+\bar{C}_0)^2 e^{\tilde{K}+K}]
      \nu_{e\Lambda}^{(1)} \nu_{e\Sigma}^{(1)} + \right. \nonumber \\
&&    \left. e^{2 \tilde{K}} \nu_{e\Lambda}^{(2)} \nu_{e\Sigma}^{(2)}
      - 4 e^{2\tilde{K}+K} (C_0+\bar{C}_0)
      (\nu_{e\Lambda}^{(1)} \nu_{e\Sigma}^{(2)} + \nu_{e\Lambda}^{(2)}
       \nu_{e\Sigma}^{(1)})\right\}.
\end{eqnarray} 
It is fairly obvious from this equation that the classical potential
will not vanish for generic moduli and non-zero $\nu_{e\Lambda}^{(i)}$
(see also section \ref{minima}).
This distinguishes this model from that of \cite{minhiggs}, for which
the classical potential vanished identically, and, for an appropriate
choice of charges, there was partial supersymmetry breaking to $N=1$.
This difference
can be understood as arising from the quaternionic structure of the manifold.
Specifically, the difference comes from the fact that in the current model,
$h_{uv}k^u_\Lambda k^v_\Sigma \neq {\cal P}^x_\Lambda {\cal P}^x_\Sigma$,
while there was equality in the model of \cite{minhiggs}.  This can be seen in
more detail by using the fact that equation~(\ref{easyP}) holds for both
models; hence, (since $\omega^x_{[u} \omega^x_{v]} = 0$)
\begin{equation} \label{notmetric}
{\cal P}^x_\Lambda {\cal P}^x_\Sigma = \omega^x_{(u} 
     \omega^x_{v)} k^u_\Lambda k^v_\Sigma.
\end{equation}
So, $\omega^x_{(u} \omega^x_{v)}$ behaves like a metric in equation
(\ref{notmetric}).  In fact, the quaternionic manifold in \cite{minhiggs}
is sufficiently trivial that $\omega^x_{(u} \omega^x_{v)}$ {\it is}  the
metric (parallel to $k^u_\Lambda$);
however, in the current case, equations (\ref{Omegas}) shows that
$\omega^x_{(u} \omega^x_{v)}$ is not even hermitian (with respect to the
complex structure defining $S$ and $C_0$ as complex variables)!  The
resultant mismatched factors in equation~(\ref{whatpot}) are then not
surprising.

\subsection{The General Case} \label{magpotsect}
The only change from the previous subsection, is that for non-zero
$\nu_m^{(i)\Lambda}$, equation~(\ref{finalterm}) gives, in addition to
equation~(\ref{killingvect}),
\begin{equation} \label{magkillingvect}
\tilde{k}^{u\Lambda} = i {\nu}^{(1)\Lambda}_m
     \left(\frac{\partial}{\partial C_0} - \frac{\partial}{\partial
           \bar{C}_0}\right)^u +
     i \nu^{(2)\Lambda}_m \left(\frac{\partial}{\partial S}
           - \frac{\partial}{\partial \bar{S}} \right)^u.
\end{equation}
As above, this gives
\begin{mathletters} \label{magps}
\begin{eqnarray}
\label{magp1}
&\tilde{{\cal P}}^{1\Lambda} = 4 e^{\frac{\tilde{K}}{2}+\frac{K}{2}}
        {\nu}^{(1)\Lambda}_m, \\
\label{magp2}
&\tilde{{\cal P}}^2_\Lambda = 0,& {\rm and} \\
\label{magp3}
&\tilde{{\cal P}}^3_\Lambda = 4 (C_0 + \bar{C}_0) e^{\tilde{K}+K}
     {\nu}^{(1)\Lambda}_m -  e^{\tilde{K}} \nu^{(2)\Lambda}_m.
\end{eqnarray}
\end{mathletters}
As discussed at the end of section~\ref{expect}, it is always
possible to choose a symplectic basis so $\nu^{(1)\Lambda}_m=0$.

The classical potential is obtained by substituting equations 
(\ref{killingvect}), (\ref{magkillingvect}), (\ref{ps}) and (\ref{magps})
into equation~(\ref{magpot}).

\section{Minima of the Electric Potential} \label{minima}
In the local case, there is no loss of generality in taking the charge
vectors all electric, and only $\nu^{(1)}_{e0}, \nu^{(2)}_{e0}$ and
$\nu^{(2)}_{e1}$ non-zero.  If $L^1$ and $L^0$ are linearly independent,
then it is straightforward to show that the potential is extremized only
when the charge vectors vanish identically.  Therefore
the case where $L^1$ and $L^0$ are linearly dependent is examined; this
is equivalent~\cite{noprepot} to demanding the non-existence of the
holomorphic prepotential for the K\"{a}hler potential.%
\footnote{Actually, Brian Greene has pointed out to me that
the prepotential always exists as one can calculate it in a basis
where the $X^\Lambda$s are linearly independent.
However, as is common in the
literature, I will use the term nonexistence to mean that in the chosen
basis, $\frac{1}{2} F_\Lambda X^\Lambda$ is not a prepotential.}

In particular, choose,
\begin{equation} \label{l0l1}
L^1 = e^{i \alpha} L^0.
\end{equation}
Equation (\ref{l0l1}) implies, via equation (\ref{defU}),
\begin{equation} \label{u00}
U^{00} = e^{-i \alpha} U^{01} = e^{i \alpha} U^{10} = U^{11}.
\end{equation}
This is consistent with the hypothesis that
\begin{equation} \label{uasl}
U^{00} = \lambda L^0 \bar{L}^0.
\end{equation}
Note that since both $U^{00}$ and $L^0 \bar{L}^0$ are positive, that
$\lambda > 0$.
It is worth noting that equations~(\ref{l0l1})-(\ref{uasl}) hold
in the $SU(1,1)/U(1)$ models of~\cite{minhiggs,higgsseq}, with
$\alpha = \frac{\pi}{2}$
and $\lambda = 1$.  In fact, equation~(\ref{l0l1}) can be derived,
in the absence of charge quantization (i.e. allowing for $Sp(h_{21}+1,\IR)$
transformations) from the linear depedence of the $L^\Lambda$s.

To look for minima of the potential on the hypermultiplet moduli space,
the variation of the potential with respect to the
dilaton multiplet is taken, and set to zero.  The variation is
(recall that in the chosen basis, all $\nu^{(i)}_{e\Lambda\geq 2} = 0$
and $\nu^{(1)}_{e1} = 0$)
\begin{eqnarray} \label{varpot}
\delta V = & L^0 \bar{L}^0 &\left\{
     16 \left((2-\lambda)-2(1+\lambda) (C_0 + \bar{C}_0)^2 e^{\tilde{K}+K}
     \right)
     e^{2 \tilde{K}+K} (\nu^{(1)}_{e0})^2
     - \right. \nonumber \\
&& \left.
     2 (1+\lambda)e^{3 \tilde{K}} |\nu^{(2)}_{e0}+e^{i\alpha}\nu^{(2)}_{e1}|^2
     + \right. \nonumber \\
&& \left. 16 (1+\lambda)(C_0+\bar{C}_0) e^{3\tilde{K}+K}
     \nu^{(1)}_{e0}
     (\nu^{(2)}_{e0}+ \nu^{(2)}_{e1}\cos \alpha)\right\}
     (\delta S+\delta \bar{S}) + \nonumber \\
& L^0 \bar{L}^0 & \left\{
     32\left((3\lambda-3) + 4(1+\lambda)(C_0+\bar{C}_0)^2e^{\tilde{K}+K}\right)
     (C_0+\bar{C}_0) e^{2\tilde{K}+2K}
     (\nu^{(1)}_{e0})^2 + \right. \nonumber \\
&& \left. 8(1+\lambda) (C_0+\bar{C}_0) e^{3\tilde{K}+K}
     |\nu^{(2)}_{e0}+e^{i\alpha}\nu^{(2)}_{e1}|^2 - \right. \nonumber \\
&& \left. 8 (1+\lambda)(1+8 (C_0+\bar{C}_0)^2 e^{2\tilde{K}+K})
     e^{2\tilde{K}+K} \nu^{(1)}_{e0}
     (\nu^{(2)}_{e0}+ \nu^{(2)}_{e1}\cos \alpha) \right\}
     (\delta C_0 + \delta \bar{C}_0),
\end{eqnarray}
and so
generically $\nu^{(i)}_{e\Lambda}=0$.  That is just the usual Calabi-Yau
compactification and so is uninteresting for this paper.
At special points on the moduli space, however, specifically those
for which
\begin{mathletters} \label{conditions}
\begin{equation} \label{cond1}
e^K = \beta e^{\tilde{K}},
\end{equation}
and
\begin{equation} \label{cond2}
(C_0+\bar{C}_0) = \frac{1}{\gamma} e^{-\tilde{K}},
\end{equation}
\end{mathletters}
where $\beta>0$ and $\gamma$ are real constants,
minimizing the potential corresponds to
solving the two equations
\begin{mathletters} \label{pot0}
\begin{eqnarray} \label{pot01}
0 &=& 8 \beta \left(\frac{2-\lambda}{1+\lambda}-2\frac{\beta}{\gamma^2}\right)
    (\nu^{(1)}_{e0})^2 -
(\nu^{(2)^2}_{e0} + \nu^{(2)^2}_{e1} + 2 \nu^{(2)}_{e0} \nu^{(2)}_{e1}
          \cos \alpha) + \nonumber \\
&& 8 \frac{\beta}{\gamma} \nu^{(1)}_{e0} (\nu^{(2)}_{e0} + 
          \nu^{(2)}_{e1} \cos \alpha), \\
\label{pot02}
0 &=& 4 \frac{\beta}{\gamma} \left(3 \frac{\lambda-1}{\lambda+1} +
          4\frac{\beta}{\gamma^2} \right)
          (\nu^{(1)}_{e0})^2 +
\frac{1}{\gamma} (\nu^{(2)^2}_{e0} + \nu^{(2)^2}_{e1} + 2 \nu^{(2)}_{e0}
         \nu^{(2)}_{e1} \cos \alpha) - \nonumber \\
&&(1+8 \frac{\beta}{\gamma^2})\nu^{(1)}_{e0} (\nu^{(2)}_{e0} +
          \nu^{(2)}_{e1} \cos \alpha).
\end{eqnarray}
\end{mathletters}
These have solutions 
\begin{mathletters} \label{minpot}
\begin{eqnarray} \label{minpot20}
&\nu^{(2)}_{e0} = 4 \frac{\beta}{\gamma} \nu^{(1)}_{e0} \pm
      2 \sqrt{\frac{2 \beta (2-\lambda)}{1+\lambda}}
      \nu^{(1)}_{e0} \cot \alpha, \\ \label{minpot21}
&\nu^{(2)}_{e1} = \mp 2 \sqrt{\frac{2 \beta (2-\lambda)}{1+\lambda}}
      \nu^{(1)}_{e0} \csc \alpha.
\end{eqnarray}
\end{mathletters}
Note that this is only well defined for $-1 < \lambda \leq 2$ (and,
as noted above, $\lambda>0$).
These can be integer valued only for special values of 
$\alpha, \beta, \gamma, \lambda$.  Also, if 
$\nu^{(1)}_{e\Lambda} = 0$, then $\nu^{(2)}_{e\Lambda}=0$.  This contradicts
the prediction made in \cite{kallosh2}; however,
this makes sense because for this set of models,
any set of $\nu^{(2)}_{e\Lambda}$s
can be symplectically transformed into a new basis in which, say,
only $\nu^{(2)}_{e0}
\neq 0$, and it was shown in \cite{joenandy} that the potential has no
minimum in this case.  Rather, to have a minimum for the potential requires
NS-NS expectation values in addition to the RR ones of \cite{joenandy} (as
predicted therein).  However, from equation~(\ref{whatisexpphi}) and
(\ref{cond1}),
the string coupling constant is
\begin{equation} \label{gen4dila}
e^{\hat{\varphi}} = \sqrt{\frac{2}{\beta}},
\end{equation}
and so $e^{\hat{\varphi}} \langle H^{(2)} \rangle$
is $O(1)$ (for small but non-zero
integral $\nu^{(1)}_{e\Lambda}$).  This contradicts the statement that
the expectation values act only as perturbations of the Calabi-Yau
compactification.  That is, the solution of equation~(\ref{minpot})
is just outside the validity of the perturbative approximation, and
so cannot be trusted.

The value of the potential at the minima of equations (\ref{minpot}) is
\begin{equation} \label{potatmin}
8 \frac{\lambda-2}{\beta} (\nu^{(1)}_{e0})^2 L^0 \bar{L}^0 e^{2 K}.
\end{equation}
This vanishes for non-zero integral charges only for $\lambda = 2$.

Also, the determinant of the matrix governing the supersymmetry transformation
of the gaugino, at the minima of the potential, is
\begin{equation} \label{detwatmin}
\det \bbox{\sf W}^a = \beta^{-1} e^{2 K} (g^{a \bar{b}}
     \bar{f}^{0}_{\bar{b}})^2 (4 + e^{2i\alpha})
     (\nu^{(1)}_{e0})^2
\end{equation}
which is never zero and so implies that the gaugino transforms under both
supersymmetries
and hence that
there are no unbroken supersymmetries.  Thus,
there is no partial supersymmetry breaking.


\section{Minima of the General Potential} \label{magminima}
In the previous section it was assumed that the symplectic section
can be chosen so that equations~(\ref{l0l1}) and
(\ref{uasl}) hold.  In addition, it will now be convenient to
make analogous assumptions for the $M_\Lambda$s; specifically it will
be assumed that
\begin{equation} \label{m0m1}
M_1 = e^{i \tilde{\alpha}}M_0
\end{equation}
and
\begin{equation} \label{maguasm}
g^{a \bar{b}} h_{a0} \bar{h}_{\bar{b}0} = \tilde{\lambda}M_0 \bar{M}_0.
\end{equation}
In addition it will be assumed that
\begin{equation} \label{mixuasls}
g^{a \bar{b}} f_a^0 \bar{h}_{\bar{b}0} = \sqrt{\lambda\tilde{\lambda}}
L^0 \bar{M}_0.
\end{equation}
These formulas are again justified by their validity
for the $SU(1,1)/U(1)$ model of \cite{minhiggs}, with $\tilde{\alpha}
=\alpha=\frac{\pi}{2}$ and $\tilde{\lambda}=\lambda=1$.

It is apparent that the general potential contains terms proportional to
$L^0 \bar{L}^0$, $L^0 \bar{M}_0$, $M_0 \bar{L}^0$ and $M_0 \bar{M}_0$
and that the coefficients of the terms proportional to
$L^0 \bar{M}_0$ and $M_0 \bar{L}^0$ are complex conjugate.  It is also
apparent that the coefficient of $L^0 \bar{L}^0$ is the same as in
the electric case, and is the same as the coefficient of $M_0 \bar{M}_0$
with $n^{(i)}_{e\Lambda}\rightarrow n^{(i)\Lambda}_m$.  So, as
$M_0$ is linearly independent of $L^0$, the minimum
of the potential is given by a subset of equations (\ref{conditions})
and two copies of equation~(\ref{minpot}), with
$n^{(i)}_{e\Lambda}\rightarrow n^{(i)\Lambda}_m$ in one of those copies.
Varying the coefficients of $L^0 \bar{M}_0$ and its complex conjugate,
leads to an inconsistency, unless either $\nu^{(i)\Lambda}_m=0$ (and/or
$\nu^{(i)}_{e\Lambda}=0$) or $\lambda=\tilde{\lambda}$.
The first case has been discussed in section~\ref{minima}; the
second then leads only to a correlation between the choice of sign in both
copies
of equation~(\ref{minpot}) (the same choice of sign must be made).  But, since
the choice of basis was such that $\nu^{(i)\Lambda}_m$=0, this means
that all magnetic charges vanish, and the situation is that of section~%
\ref{minima}.

\section{Singularities} \label{conifold}
Points where $X^\Lambda$ and/or $F_\Lambda$ vanish are conifold
singularities.  At these points the theory appears to be singular,
but this is because black holes become massless at these points, and
so need to be ``integrated in.''~\cite{coniandy,coni}
The 10-dimensional description of the black holes is that of 3-branes wrapped
around Calabi-Yau 3-cycles.  The black holes are massless when the 3-cycle
volume vanishes.  These
conifold singularities
appear on complex codimension one surfaces in the moduli space.  The black
holes have unit charge with respect to the $U(1)$ and electric or magnetic,
corresponding to the
vanishing period of the degenerating 3-cycle.
($(X^\Lambda, F_\Lambda)$ is the symplectic
period vector.)
More complicated singularities
(which are called Argyres-Douglas points in this
paper) occur on complex codimension two surfaces
where two surfaces on which there are conifold singularities intersect.
These singularities were first discovered in a field-theoretic context
in \cite{adpts} and their relevance to string theory was
given in \cite{greenes}.
At these points two black holes become massless and their charge vectors
can be non-local.

A massless black hole is included in the low energy theory as
a hypermultiplet.  As in~\cite{joenandy}, only black holes with the
same types of charges as the dilaton, will be considered.
Of course, even with this restriction, not all sets of charges
will be physically realizable at conifold and/or Argyres-Douglas points, but this
will not affect the discussion.  Also, since the black hole charges are
associated with the vanishing periods, equation~(\ref{magmass}) shows
that the gravitino mass matrix is continuous.  Similarly, the matrix
governing the hyperino
variation is continuous at the singularities (see equation~(\ref{magmatrixN}));
however, the gaugino variation matrix, equation~(\ref{magmatrixW}) is not
continuous at a singularity, because the appropriate
$f^\Lambda_a$s and $h_{a\Lambda}$s
will not vanish there.

In~\cite{joenandy}, it was shown that on the IIA side with only
10-dimensional RR Calabi-Yau expectation values, there are
flat directions with $N=2$ supersymmetry at conifold points.  These
flat directions were those for which it was possible to set
\begin{equation} \label{joenandyN2}
{\cal P}^x_\Lambda = 0.
\end{equation}
A similar result will be shown here, in the more general case of both
NS-NS and RR expectation values, and also at Argyres-Douglas points.

First consider Argyres-Douglas points.  The black hole hypermultiplets are each doublets,
$\bbox{{\sf B}_1}$ and $\bbox{{\sf B}_2}$.
At least to lowest order in an expansion
in $\bbox{{\sf B}_1}$ and $\bbox{{\sf B}_2}$, and about the singularity,
each element of the black hole doublet has the same charge, so the
Killing vectors, equations~(\ref{killingvect}) and (\ref{magkillingvect})
become
\begin{mathletters} \label{bhkillingvects}
\begin{eqnarray} \label{bhkillingvect0}
k^u_0 &=& i {\nu}^{(1)}_{e0}
     \left(\frac{\partial}{\partial C_0} - \frac{\partial}{\partial
           \bar{C}_0}\right)^u +
     i \nu^{(2)}_{e0} \left(\frac{\partial}{\partial S}
           - \frac{\partial}{\partial \bar{S}} \right)^u + \nonumber \\
&&   i \left(\bbox{{\sf B}^{\sf T}_1} \frac{\partial}{\partial
           \bbox{{\sf B}_1}} -
           \bbox{{\sf B}^\dagger_1} \frac{\partial}{\partial
           \bbox{{\sf \bar{B}}_1}} \right) +
     i n^{(2)}_{Be0} \left(\bbox{{\sf B}^{\sf T}_2}
           \frac{\partial}{\partial \bbox{{\sf B}_2}} -
           \bbox{{\sf B}^\dagger_2} \frac{\partial}{\partial
           \bbox{{\sf \bar{B}}_2}} \right), \\
\label{bhkillingvect1}
k^u_1 &=& i \nu^{(2)}_{e1} \left(\frac{\partial}{\partial S}
           - \frac{\partial}{\partial \bar{S}} \right)^u +
     i n^{(2)}_{Be1}
           \left( \bbox{{\sf B}^{\sf T}_2} \frac{\partial}{\partial
           \bbox{{\sf B}_2}} -
           \bbox{{\sf B}^\dagger_2} \frac{\partial}{\partial
           \bbox{{\sf \bar{B}}_2}} \right),\\
\label{bhmagkillingvect0}
\tilde{k}^{0u} &=& i \nu^{(2)0}_{m} \left(\frac{\partial}{\partial S}
           - \frac{\partial}{\partial \bar{S}} \right)^u +
     i n^{(2)}_{Bm} \left( \bbox{{\sf B}^{\sf T}_2} \frac{\partial}{\partial
           \bbox{{\sf B}_2}} -
           \bbox{{\sf B}^\dagger_2} \frac{\partial}{\partial
           \bbox{{\sf \bar{B}}_2}} \right),
\end{eqnarray}
\end{mathletters}
where $(n^{(i)\Lambda}_m, n^{(i)}_{e\Lambda})$ are the black hole
charge vectors.
To lowest order, the $SU(2)$ connection on the black hole
quaternionic manifold can be ignored, and the triplet of K\"{a}hler forms
can be taken to be $\Omega^x = -i d\bbox{{\sf B}_i^\dagger}\wedge \sigma^x
d\bbox{{\sf B}_i}$, giving
\begin{mathletters} \label{bhPxs}
\begin{eqnarray} \label{bhPx}
{\cal P}^x_\Lambda &=& {\cal P}^x_\Lambda |_{\bbox{{\sf B}_i} = \bbox{0}}
      + \bbox{{\sf B}^\dagger_1} \sigma^x \bbox{{\sf B}_1} \delta^0_\Lambda
      + n^{(2)}_{Be\Lambda}
        \bbox{{\sf B}^\dagger_2} \sigma^x \bbox{{\sf B}_2}, \\
\label{magbhPx}
\tilde{\cal P}^{0x} &=& \tilde{\cal P}^{0x} |_{\bbox{{\sf B}_i} = \bbox{0}}
      + n^{(2)}_{Bm} \bbox{{\sf B}^\dagger_2} \sigma^x \bbox{{\sf B}_2}.
\end{eqnarray}
\end{mathletters}
where ${\cal P}^x_\Lambda |_{\bbox{{\sf B}_i} = \bbox{0}}$ and
$\tilde{\cal P}^{0x} |_{\bbox{{\sf B}_i} = \bbox{0}}$ were given
in equations~(\ref{ps}) and (\ref{magps}).  Taking $L^0$ and $M^0$ to
be the vanishing periods in a symplectic basis where $L^0$ and $L^1$ are
linearly independent, it can be shown
that the hyperino variation cannot have any
zero eigenvectors, and hence that there will be no supersymmetric
minima of the potential.  (This analysis requires the quaternionic
vielbein which was given in~\cite{bodcadb} and is essentially
equation (\ref{whate}).)  If there is linear dependence,
then $L^1$ also vanishes, and so the gravitino mass matrix vanishes.
Then, the classical potential is non-negative, so the only vacua have
non-negative cosmological constant; i.e. the vacua are (asymptotically)
flat or de Sitter.  Since de Sitter spaces do not admit a supersymmetry
algebra,
any supersymmetric minimum of the potential (if such a minimum exists) must
occur at points where the potential vanishes.  These minima have $N=2$
supersymmetry since for non-positive cosmological constant the number
of supersymmetries is equal to the number of massless gravitini~%
\cite{wardid1}.  This requires
\begin{equation} \label{admin}
{\cal P}^x_\Lambda = \tilde{\cal P}^{0x} = 0.
\end{equation}
These have solutions when, for example, the black hole charges are
proportional to the dilaton charges.  This result has also been obtained
via explicit calculation.

In that case, equations (\ref{admin}) are six equations in eight (real)
unknowns.  Thus the flat directions are parametrized by two real numbers,
which correspond to the overall phase of the hypermultiplets.  These are
the would-be goldstone bosons that are eaten by the vectors; as in~%
\cite{coni}, there is a transition at the Argyres-Douglas points from a Calabi-Yau
compactification with Hodge numbers $(h_{11},h_{21})$ to one with
Hodge numbers $(h_{11}, h_{21}-2)$.

The above discussion also holds for conifold points by taking either
$\bbox{{\sf B}_1}=\bbox{0}$ or $\bbox{{\sf B}_2}=\bbox{0}$, in addition
to the above.  Again, equation (\ref{admin}) may not have solutions for
all choices of charge vectors.  Of course, this time only one black-hole
hypermultiplet is being eaten by a vector multiplet, so the transition is
to a Calabi-Yau with Hodge numbers $(h_{11},h_{21}-1)$.

\section{Conclusion} \label{conc}
When Type IIB strings are compactified on a
Calabi-Yau manifold (with $h_{21} \geq 1$) and Calabi-Yau valued
expectation values are given to the NS-NS and RR 3-forms, the dilaton
is given electric and magnetic charges.  The classical potential was
derived in this situation.  Under the assumption that the special
K\"{a}hler moduli space
of complex structures of the Calabi-Yau has a symplectic basis for which
there is no prepotential (and some auxiliary assumptions, most of which
would be unnecessary if $Sp(h_{21}+1,\IR)$
transformations were allowed instead of just
$Sp(h_{11}+1,\IZ)$) it was shown that
for certain values of the charges, the
potential could be minimized, though not while remaining within the
validity of the calculation.  $N=2$ supersymmetric minima are obtained
at conifold points, Argyres-Douglas points and, as in~\cite{joenandy}, in
the infinite Calabi-Yau volume limit.
It is interesting that the $N=0$ minima are
below the $N=2$ minima.  In fact, from equations~(\ref{potatmin}),
(\ref{conditions}), (\ref{whatK}) and (\ref{defS}),
it is seen that the global minimum
of the potential
($V \rightarrow -\infty$) occurs in the limit of vanishing Calabi-Yau
volume.  It has been shown in~\cite{hullpet} that $N=0$ vacua are
classically stable if they occur at global minima of the potential.
Unfortunately, it is neither clear that this would hold quantum mechanically,
nor likely, since the vanishing Calabi-Yau volume limit is both well outside
the limit of validity of the calculation and well inside the region where
significant quantum and stringy effects are expected.

It was found that partial supersymmetry breaking cannot occur.  This agrees
with~\cite{cand} where the conditions for Type IIB compactified to $D=4$,
to have supersymmetry were found and it was discovered that there was
$N=2$ or $N=0$.  This problem was also studied in~\cite{dewitIIB}, with
a warp factor (Calabi-Yau-valued conformal factor for the space-time
metric) included, but with the same conclusion.  This remains true at
singularities in the moduli space.

\acknowledgments
I am grateful to Katrin Becker, Melanie Becker, Sergio Ferrara,
David Kaplan, Tom\'{a}s Ort\'{\i}n, John Pierre, Joe Polchinski, and 
Andrew Strominger for many useful discussions.
I am especially indebted to Andrew Strominger for critical
readings of the manuscript.
I thank NSERC and NSF for financial support.
This work was also partially supported by DOE Grant No. DOE-91ER40618.

\appendix
\section*{Compactification of IIB on a Calabi-Yau}
In this appendix, the compactification of type IIB supergravity on a
Calabi-Yau manifold is discussed, following \cite{bodcadb}.
Therefore instead of using the non-self-dual
action of equation~(\ref{nsdlag}), the
type IIB equations of motion \cite{schwarz} will be used.
Also, as
in \cite{bodcadb}, attention is restricted to an $h_{11}=1, h_{21}=0$
Calabi-Yau.  The (uncomplexified)
moduli space therefore is one-dimensional, and
corresponds to the choice of metric; specifically a conformal factor~%
$e^\sigma$.  Furthermore, as RR fields are suppressed in string
perturbation theory, and because only the structure
of the dilaton multiplet is of interest, it will be convenient to
take\footnote{It is interesting that if the two and four form
field strengths are not assumed to vanish on the Calabi-Yau, then for a
Calabi-Yau with $h_{11}>1$, the fact that the wedge product of two
harmonic $(1,1)$-forms is not harmonic, means that the ansatz for the
four form necessarily involves non-harmonic three-forms.
Nevertheless, (and fortunately),
it turns out that no residual effects of the three-forms
appear in the four-dimensional action.}

\begin{equation} \label{assumption}
\hat{l} = 0; \hat{B}_{i\bar{j}} = 0;
D_{\hat{\mu}\hat{\nu}\hat{\sigma}
     \hat{\tau}}=0.
\end{equation}
The self-duality of the five-form field strength is then devoid of
content\cite{bodcadb}.  (This is not inconsistent with equation~(\ref{foncy})
since the vectors do not mix with the scalars and only the scalars are
being considered here.)

The equations of motion are usually written in terms of the
fields\cite{schwarz,fields}
\begin{mathletters} \label{defeomfields}
\begin{eqnarray}
\label{defpsi}
&&\hat{\psi} = \frac{1+i\hat{\lambda}}{1-i\hat{\lambda}} =
       \frac{1-e^{-\hat{\varphi}}}{1+e^{-\hat{\varphi}}}, \\
\label{defp}
&&\hat{P}_{\hat{\mu}} = \frac{\partial_{\hat{\mu}} \hat{\psi}}
       {1-\hat{\psi}^*\hat{\psi}}, \\
\label{defq}
&&\hat{Q}_{\hat{\mu}} = \frac{{\rm Im}(\hat{\psi} \hat{\partial}_{\hat{\mu}}
       \hat{\psi}^* )}{1-\hat{\psi}^*\hat{\psi}}, \\
\label{defg}
&&\hat{G}_{\hat{\mu}\hat{\nu}\hat{\rho}} = \frac{\hat{H}_{\hat{\mu}\hat{\nu}
       \hat{\rho}} - \hat{\psi} \hat{H}^*_{\hat{\mu}\hat{\nu}\hat{\rho}}}
       {(1-\hat{\psi}^*\hat{\psi})^{\frac{1}{2}}};
       \hat{H}_{\hat{\mu}\hat{\nu}\hat{\rho}} =
       \hat{H}_{\hat{\mu}\hat{\nu}\hat{\rho}}^{(1)} +
       i \hat{H}_{\hat{\mu}\hat{\nu}\hat{\rho}}^{(2)}. 
\end{eqnarray} 
\end{mathletters}
The equation of motion that will be most
interesting is
\begin{equation} \label{eom}
(\nabla_{\hat{\mu}} - i\hat{Q}_{\hat{\mu}})
     \hat{G}^{\hat{\mu}}{}_{\hat{\nu}\hat{\rho}} = \hat{P}_{\hat{\mu}}
     \hat{G}^{*\hat{\mu}}{}_{\hat{\nu}\hat{\rho}}.
\end{equation}
Equation~(\ref{eom}) is satisfied trivially on the Calabi-Yau.  After
performing a 4-dimensional Weyl rescaling $g_{\mu \nu} \rightarrow
e^{-3 \sigma} g_{\mu \nu}$, equation~(\ref{eom}) becomes (on the
spacetime)\footnote{Note that this differs slightly from equation~(2.18)
of \cite{bodcadb}; however, equations~(\ref{finaleom}) and (\ref{defD})
agree with \cite{bodcadb}.}
\begin{eqnarray} \label{messyeom}
(\nabla_\mu + &&\frac{\psi^*\partial_\mu\psi-\psi\partial_\mu\psi^*}
     {2(1-\psi^*\psi)})e^{3 \sigma} G^{\mu}{}_{\nu \rho} \nonumber \\
&& = \frac{\partial_{\mu}\psi}{1-\psi^* \psi}e^{3\sigma} G^{*\mu}{}
     _{\nu\rho}.
\end{eqnarray}
Subtracting $\psi$ times the complex conjugate of equation~(\ref{messyeom}),
from equation~(\ref{messyeom}), gives
\begin{equation} \label{finaleom}
(1-\psi^*\psi)^{\frac{1}{2}} \nabla_\mu \left[ e^{3\sigma}
     \frac{G^{\mu}{}_{\nu \rho}-\psi G^{*\mu}{}_{\nu \rho}}
     {(1-\psi^*\psi)^{\frac{1}{2}}} \right],
\end{equation}
which is satisfied by introducing a complex scalar field $D$ such that
\begin{equation} \label{defD}
\partial_\mu D = e^{3 \sigma} \frac{G_{\mu}-\psi G^*_\mu}
     {(1-\psi^*\psi)^{\frac{1}{2}}},
\end{equation}
where, as in equation~(\ref{defhdual}),
\begin{equation} \label{defgdual}
G_{\mu \nu \rho} = \varepsilon_{\mu \nu \rho}{}^\sigma G_\sigma.
\end{equation}

The other equation of motion that is used in \cite{bodcadb} is
\begin{eqnarray} \label{einseqn}
\hat{R}_{\hat{\mu} \hat{\nu}} = 2\hat{P}_{(\hat{\mu}} \hat{P}^*_{\hat{\nu})}
     + \frac{9}{4} \hat{G}_{(\hat{\mu}}{}^{\hat{\sigma}\hat{\tau}}
     \hat{G}^*_{\hat{\nu})
     \hat{\sigma}\hat{\tau}} - \frac{3}{16} \hat{g}_{\hat{\mu} \hat{\nu}}
     \hat{G}^{\hat{\nu}\hat{\sigma}\hat{\tau}}
     \hat{G}^*_{\hat{\nu}\hat{\sigma}\hat{\tau}}.
\end{eqnarray}
By substituting the Calabi-Yau part of this equation
into the space-time part of the equation, the four-dimensional
action
\begin{eqnarray} \label{triviallag}
S = \int d^4x \sqrt{-g} \left\{ \frac{1}{2} R + |P_\mu|^2 + 3 (\partial_\mu
     \sigma)^2 + \frac{9}{4} |G_\mu|^2 \right \}
\end{eqnarray}
can be deduced.\footnote{This equation differs from the corresponding
formula in \cite{bodcadb} in an essential way.}
Alternatively, this can be found, almost by inspection,
via dimensional reduction of the NSD action of equation~(\ref{nsdlag}).
As mentioned above, there is a space-time dependent conformal factor of
$e^\sigma$ in the Calabi-Yau metric; hence
$\sqrt{-\hat{g}}=e^{3\sigma} \sqrt{-g}$ and so to remain in the Einstein
frame required the Weyl rescaling of the four dimensional metric $g_{\mu
\nu} \rightarrow e^{-3 \sigma} g_{\mu \nu}$.  This is the same Weyl
rescaling used in the derivation of equation~(\ref{finaleom}) and the
reason for it.

To obtain the standard quaternionic geometry, make the field
redefinitions,\footnote{Equation~(\ref{defZ}) differs 
from the corresponding formula in 
\cite{bodcadb} in an essential way.}
\begin{mathletters} \label{defs}
\begin{eqnarray} 
\label{defZ}
Z = -i e^{\sigma+\frac{1}{2}\varphi}; \\
\label{defC0}
C_0 = i \frac{3 \sqrt{2}}{4} {\rm Im} D; \\
\label{defphi}
\phi = e^{3 \sigma-\frac{1}{2}\varphi}; \\
\label{defphitil}
\tilde{\phi} = {\rm Re} D; & {\rm and} \\ 
\label{defS}
S = \phi + i \tilde{\phi}.
\end{eqnarray}
\end{mathletters}
Define also 
\begin{mathletters} \label{moredefs}
\begin{eqnarray}
\label{defK}
K = -\ln(-\frac{i}{8}(Z-\bar{Z})); \\
\label{defKtil}
\tilde{K} = -\ln[S+\bar{S}]; \\
\label{defN00}
{\cal N}_{00} = {\cal R}_{00} = \frac{i}{32} (Z-\bar{Z})^3; \\
\label{defdc}
D_\mu C_0 = \partial_\mu C_0; && {\rm and}\\
\label{defds}
D_\mu S = \partial_\mu S
\end{eqnarray}
\end{mathletters}
Then, the scalar part of the action of equation~(\ref{triviallag}) becomes
(as in \cite{fersab})
\begin{eqnarray}
S = \int d^4x \sqrt{-g} &&\left\{
     K_{Z \bar{Z}}\partial_\mu Z \partial^\mu \bar{Z} +
     \tilde{K}_{S\bar{S}}D_\mu S D^\mu \bar{S} +
     \tilde{K}_{S\bar{C}_0}D_\mu S D^\mu \bar{C}_0 + \right. \nonumber \\
&& \left. \tilde{K}_{C_0\bar{S}}D_\mu C_0 D^\mu \bar{S} +
     \tilde{K}_{C_0\bar{C}_0}D_\mu C_0 D^\mu \bar{C}_0 \right \},
\end{eqnarray}
where the subscripts on $K$, $\tilde{K}$ denote differentiation.
Note that, as defined above, $C_0$ is pure imaginary; this however, is a
consequence only of the simplifying assumptions made above and is,
of course, not general, and is not assumed in the main body of the paper.

Combining equations (\ref{defg}), (\ref{defgdual}),
(\ref{defD}), (\ref{defs})
and (\ref{moredefs})
gives equations (\ref{whatishdual}).
Also, the Weyl rescaling used here can be
reexpressed in terms of $K$ and $\phi$; this is the Weyl rescaling used
in equation~(\ref{interaction}).  These results can also be obtained from
the slightly more general formulas of \cite{bodcadb} (after the above 
corrections have been made) by keeping only terms of lowest order in the
string coupling constant $e^{\hat{\varphi}}$.

\end{document}